\documentclass[twocolumn, prl, showpacs, aps]{revtex4-1}

\usepackage{graphicx}
\usepackage{ae,aecompl}
\usepackage{bm}
\usepackage{textcomp}
\usepackage{xcolor}

\begin{document}


\date{\today}
\title{Scalable Spin Squeezing for Quantum-Enhanced Magnetometry\\ with Bose-Einstein Condensates}

\author{W. Muessel}
\email{ScalableSqueezing@matterwave.de}
\author{H. Strobel}
\author{D. Linnemann}
\author{D. B. Hume}
\author{M. K. Oberthaler}

\affiliation{Kirchhoff-Institut f\"ur Physik, Universit\"at Heidelberg, Im Neuenheimer Feld 227, 69120 Heidelberg, Germany.}

\begin{abstract}
A major challenge in quantum metrology is the generation of entangled states with macroscopic atom number. Here, we demonstrate experimentally that atomic squeezing generated via non-linear dynamics in Bose Einstein condensates, combined with suitable trap geometries, allows scaling to large ensemble sizes. We achieve a suppression of fluctuations  by $5.3(5)$\,dB for 12300 particles, which implies that similar squeezing can be achieved for more than $10^7$ atoms. With this resource, we demonstrate quantum-enhanced magnetometry by swapping the squeezed state to magnetically sensitive hyperfine levels that have negligible nonlinearity. We find a quantum-enhanced single-shot sensitivity of $310(47)$\,pT for static magnetic fields in a probe volume as small as 90\,\textmu m${^3}$.

\end{abstract}
\maketitle 


 Atom interferometry~\cite{CroninRMP2009} is a powerful technique for the precise measurement of quantities such as acceleration, rotation and frequency~\cite{GeigerNATCOMM2011,PetersMETROLOGIA2001, GustavsonPRL1997, WynandsMETROLOGIA2005}.  Since state-of-the-art atom interferometers already operate at the classical limit for phase precision, given by the projection noise $\Delta \theta_{\text{cl}} = 1/\sqrt{N}$  for $N$ detected particles~\cite{ItanoPRA1993}, quantum entangled input states are a viable route for further improving the sensitivity of these devices. \\  
One class of such states are spin squeezed states, which outperform the classical limit at a level given by the metrological spin squeezing parameter $\xi_{\text{R}}$ with $\Delta \theta_{\text{sq}} = \xi_{\text{R}} \cdot \Delta \theta_{\text{cl}} $~\cite{WinelandPRA1994, GiovannettiSCIENCE2004}.  In the photonic case, quantum-enhanced interferometry with squeezed states is routinely employed in optical gravitational wave detectors~\cite{LigoNatPhys2011}. For atoms, proof-of-principle experiments have shown that spin squeezed states can be generated in systems ranging from high-temperature vapors to ultracold Bose-Einstein condensates (BEC)~\cite{EsteveNATURE2008, AppelPNAS2009, GrossNATURE2010, RiedelNATURE2010, LerouxPRL2010-CS, ChenPRL2011, SewellPRL2012, BerradaNATCOMM2013}, surpassing the classical limit in atom interferometry~\cite{GrossNATURE2010, SewellPRL2012, OckeloenPRL2013} and atomic clocks~\cite{LouchetChauvetNJP2010,LerouxPRL2010}. \\
\begin{figure}[hb!]
\includegraphics[width = 86mm]{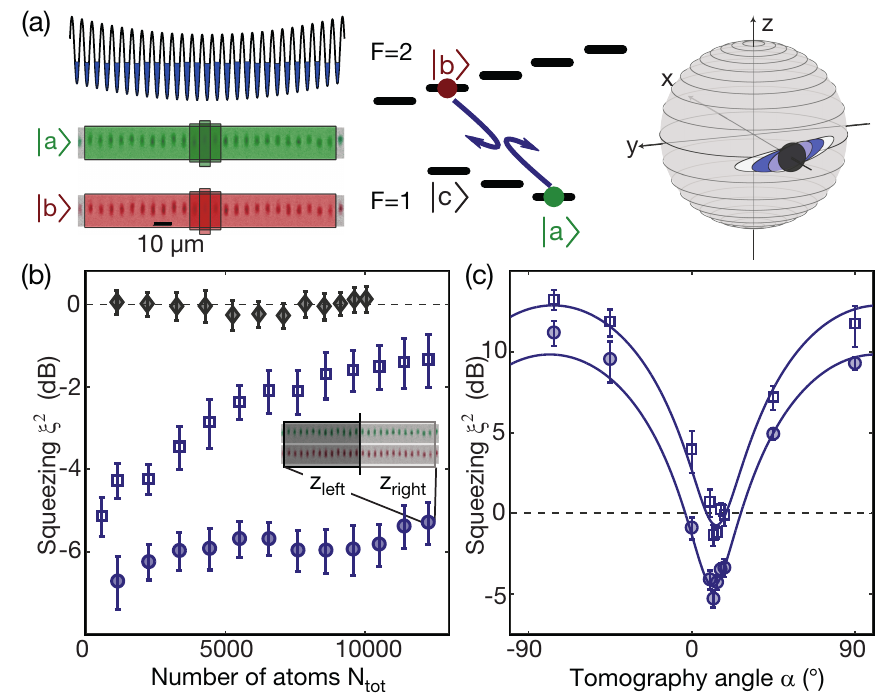}
\caption{(color online). {\bf Scaling squeezing to large ensemble sizes.} (a) 
Independent squeezing (displayed on generalized Bloch sphere) of 25 binary BECs in a 1D lattice gives access to large atom numbers by summing over adjacent lattice sites. (b) Scaling of the squeezed state after 20\,ms of nonlinear evolution. A relative analysis using adjacent parts of the lattice (inset) yields $\xi^2_{\text{rel}}  = -5.3(5)$\,dB for the full sample of 12300 atoms (blue circles), compared to $\xi^2_{\text{N}} = -1.3(6) $\,dB  using direct analysis (open squares). The corresponding measurement for an initial coherent spin state (black diamonds) is at the classical shot noise limit.  (c) Analysis for different tomography rotation angles $\alpha$ and $12300$ atoms  shows the sinusoidal behavior from the redistributed uncertainties for direct (open squares) and relative analysis (filled circles). Error bars are the statistical 1\,s.d. error.}
\label{Fig1}
\end{figure}
BECs are particularly well-suited for applications that require long interrogation times, high spatial resolution or control of motional degrees of freedom, such as in measurements of acceleration and rotation, due to their high phase-space density. However, in these systems scaling of the squeezed states to large particle numbers is intrinsically limited by density dependent losses, e.g. due to molecule formation. Keeping these processes negligible implies that for larger numbers the volume has to be increased, which in turn limits the generation of squeezed states due to uncontrolled nonlinear multimode dynamics. Here, we show how this limitation can be overcome by realizing an array of many individual condensates with an optical lattice, increasing the local trap frequencies but keeping the density small.\\
In our experiment, we simultaneously prepare up to 30 independent BECs by superimposing a deep 1D optical lattice (period 5.5\,\textmu m) on a harmonic trap with large aspect ratio, which provides transverse confinement (see Fig. 1(a)).  Each lattice site contains a condensate with  $N = 300$ to 600 atoms in a localized spatial mode and the internal state $|\text{a} \rangle = |F, m_F\rangle =|1, 1\rangle$ of the lowest hyperfine manifold. Using a two-photon radio frequency and microwave transition, we apply phase and amplitude controlled coupling of the states $|\text{a} \rangle$ and $|\text{b} \rangle=|2, -1 \rangle$, forming an effective two-level system.  A magnetic bias field of 9.12\,G brings the system near a Feshbach resonance, changing the interspecies interaction and leading to the nonlinearity necessary for squeezed state preparation. The nonlinear evolution is governed by the one-axis twisting Hamiltonian  $\mathcal{H} = \chi \hat{J}_z^2$ \cite{KitagawaPRA1993}, where the interaction strength is parametrized  by $\chi$, and  $\hat{J}_z = (\hat{N}_b-\hat{N}_a)/2$  is the z component of the Schwinger pseudospin. As indicated in  Fig. 1(a), the evolution of an initial coherent spin state under this Hamiltonian leads to an elongated squeezed state with reduced quantum uncertainty along one direction. Details about the experimental sequence can be found in the Supplementary Information \cite{SuppInfo}. \\
In order to get access to the axis of minimal fluctuations, we rotate the state by an angle $\alpha$ around its mean  spin direction. A projective measurement of the population imbalance $z= 2 J_z/N = (N_b-N_a)/N$ is implemented by state-selective absorption imaging (see Fig. 1(a)). The lower quantum uncertainty of the state translates into reduced fluctuations of $z$ for repeated experiments, and is quantified using the number squeezing parameter $\xi^2_{\text{N}} = N\cdot \text{Var}(z)$ \cite{SuppInfo}.\\
\begin{figure}
\includegraphics[width = 86mm]{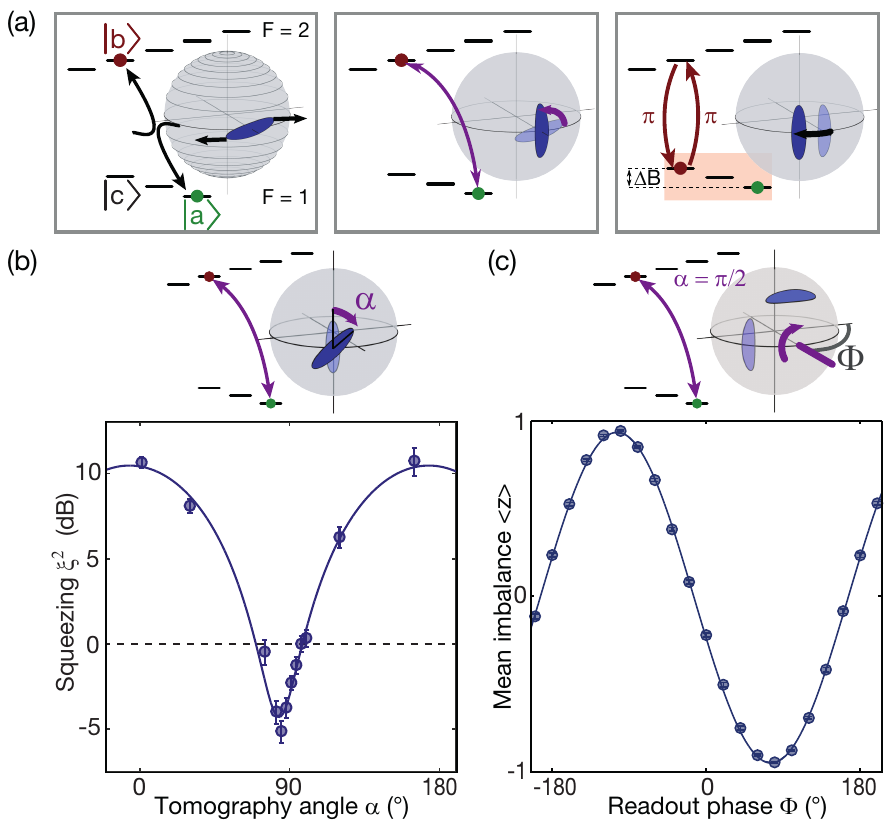}
\caption{(color online). {\bf Swapping  squeezed states for a quantum-enhanced Ramsey scheme.} (a)  After 20\,ms of nonlinear evolution on all lattice sites (left panel), the squeezed states are rotated to the phase sensitive axis (middle panel). 
By swapping the population from state $|\text{b}\rangle$ to $|\text{c}\rangle$, the nonlinearity is effectively switched off and the magnetic field sensitivity is significantly enhanced (right panel). After magnetic field-dependent phase evolution, the population is swapped back before readout. (b) Squeezing tomography after a hold time of 1\,\textmu s (negligible magnetic field phase) confirms relative squeezing of   $\xi^2_{\text{rel}} = -5.1 ^{+0.6}_{-0.7}$\,dB  ($-3.8(5)$\,dB with detection noise)  after swapping and readout. (c) Ramsey fringe for a final $\pi$/2-pulse with variable readout phase $\Phi$ yields a visibility of  $\mathcal{V} = 0.950(5) $. This implies metrologically relevant spin squeezing of  $-3.4(5)$\,dB and direct applicability of the system for quantum-enhanced gradiometry. Error bars are the statistical 1\,s.d. error. }
\label{Fig2}
\end{figure}
High-resolution imaging of the individual lattice sites allows us to study $\xi^2_{\text{N}}$ for different system sizes by summing the populations of several sites (see Fig. 1(b)). We develop a relative squeezing analysis that is insensitive to the magnetic field fluctuations present in our system (\textpm 45 \textmu G for several days). For that, we divide the lattice in half and evaluate the difference of the population imbalances $\delta z = z_{\text{left}} - z_{\text{right}}$ of the two regions (see inset Fig. 1(b)), which rejects common mode fluctuations. The corresponding squeezing parameter is given by $\xi^2_{\text{rel}}= N_{\text{tot}}/4 \cdot\text{Var}(\delta z)$ for equal particle numbers on both sides and $\langle z_i \rangle \approx$ 0  ~\cite{SuppInfo}, which is directly connected to the quantum enhancement of gradiometry, as described below. We find $\xi^2_{\text{rel}}  = -5.3(5)$\,dB for the full ensemble of 12300\,atoms after 20\,ms of nonlinear evolution (Fig. 1(b), blue circles). The remaining decrease of squeezing for large $N_{\text{tot}}$ is due to the atom number dependent parameters of the single-site Hamiltonian. This affects the squeezing as well as the optimal tomography angles for different ensemble sizes.
From our observations we infer that by extending our one-dimensional array (30 lattice sites) to three dimensions ($30^3$ sites), ensembles as large as $10^7$ atoms can be squeezed to the same level.\\ 
To compare the scaling of our squeezed state with the best attainable classical state, we show the  values of  $\xi^2_{\text{rel}}$ obtained for the initial coherent spin state (black diamonds), yielding the expected classical shot noise limit.  In the case of squeezed states even the direct analysis of summing all ensembles, which does not reject technical fluctuations, yields squeezing of  $\xi^2_{\text{N}} = -1.3(6)$\,dB for 12300 particles (Fig. 1(b), open blue squares). For all given variances, the independently characterized photon shot noise of the detection process was subtracted.\\
Fig. 1(c) shows the tomographic characterization of fluctuations for different readout rotation angles $\alpha$ for the ensemble containing $10^4$\,particles, revealing the expected sinusoidal behavior. The difference between the relative and the direct analysis is consistent with an independent characterization of the technical noise ~\cite{SuppInfo} and can be further reduced with an optimized spin-echo pulse.\\
A natural application of atomic squeezed states is the measurement of magnetic fields, where BECs are an ideal system to achieve both high sensitivity and spatial resolution \cite{VengalattorePRL2007, AignerSCIENCE2008, OckeloenPRL2013}. We implement a quantum-enhanced magnetometer using a modified Ramsey sequence that coherently transfers the population of one  level to a different hyperfine state for the interrogation time. The advantage of this state swapping is twofold;  the  nonlinear interaction becomes negligible on our interferometric timescales and the magnetic sensitivity is significantly increased from   $\approx 1$\,Hz/\textmu T (second order Zeeman shift at the operating field of $B_0 = 9.12$\,G) to $\mathcal{S} \approx  140$\,Hz/\textmu T (first order Zeeman shift). Fig. 2(a) depicts the implemented experimental sequence. After generating the squeezed state in the levels $|\text{a} \rangle$ and  $|\text{b} \rangle$ (left panel), we rotate it for maximum phase sensitivity (middle panel). The interrogation time $ t_{\text{int}}$  of the Ramsey sequence starts with a microwave $\pi$-pulse ($t_{\pi}$) which swaps the level $|\text{b} \rangle$  of the phase squeezed state to the level  $|\text{c} \rangle = |1, -1 \rangle$, yielding increased magnetic sensitivity (right panel). After a hold time $t_{\text{hold}}$, we swap the state back to the original level.
During this sequence, the state acquires a phase $\varphi = 2\pi \left(\mathcal{S}\cdot (B-B_0) + \delta\right)  \cdot ( t_{\text{hold}} + 2t_{\pi})$,  where  $\delta$  describes the  relative  detuning of the $\pi$-pulse. A Ramsey fringe is obtained by a final $\pi/2$ rotation with varied pulse phase $\phi$.\\
We first confirm that the level of squeezing is maintained during state swapping by performing an interferometric sequence with $t_{\text{hold}} = 1$\,\textmu s followed by a tomographic analysis (Fig. 2(b)). We find $\xi^2_{\text{rel}} = -5.1 ^{+0.6}_{-0.7}$\,dB at the optimum tomography angle and a Ramsey fringe visibility of $\mathcal{V} = 0.950(5)$ (Fig. 2(c)), revealing no significant reduction of the squeezing initially present. Without subtraction of detection noise, we find squeezing of $ -3.8(5)$\,dB, which corresponds to metrologically relevant spin squeezing of $-3.4(5)$\,dB  for  the visibility $\mathcal{V} = 0.95$.\\
\begin{figure}[hb]
\includegraphics[width = 86mm]{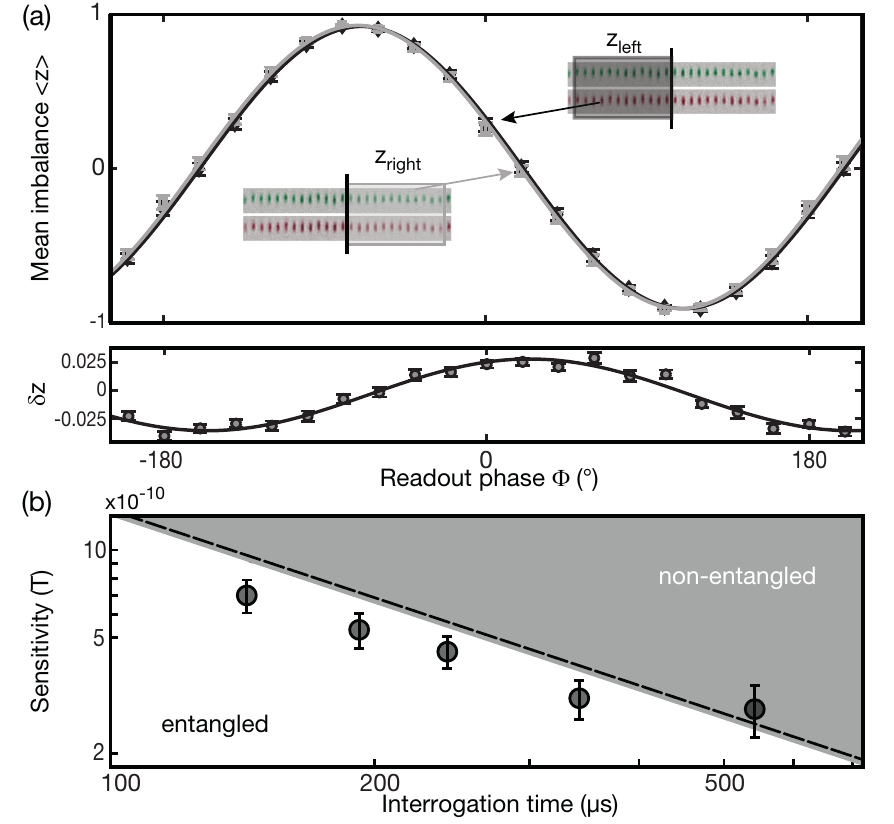}
\caption{(color online). {\bf Ramsey magnetometry beyond the standard quantum limit.} (a) A magnetic field gradient translates into a phase shift of the corresponding Ramsey fringes for left and right parts of the full sample (upper panel for $t_{\text{int}} = 342$\,\textmu s). The difference in population imbalance $\delta z =  z_{\text{left}}-z_{\text{right}}$ is maximal at the zero crossings of the individual fringes (lower panel). (b)  Magnetic field sensitivity of the Ramsey magnetometer versus interrogation time. For the full sample, we find quantum-enhanced  performance up to interrogation times of 342\,\textmu s with a single shot sensitivity of $310(47)$\,pT for static magnetic fields. For longer times, we are limited by fluctuations of the magnetic field. The shaded area depicts the region which is accessible for non-entangled states, the dashed line indicates the standard quantum limit including detection noise. Error bars show the statistical 1\,s.d. confidence intervals obtained from a resampling method.}
\label{Fig3}
\end{figure}
\begin{figure}
\includegraphics[width = 86mm]{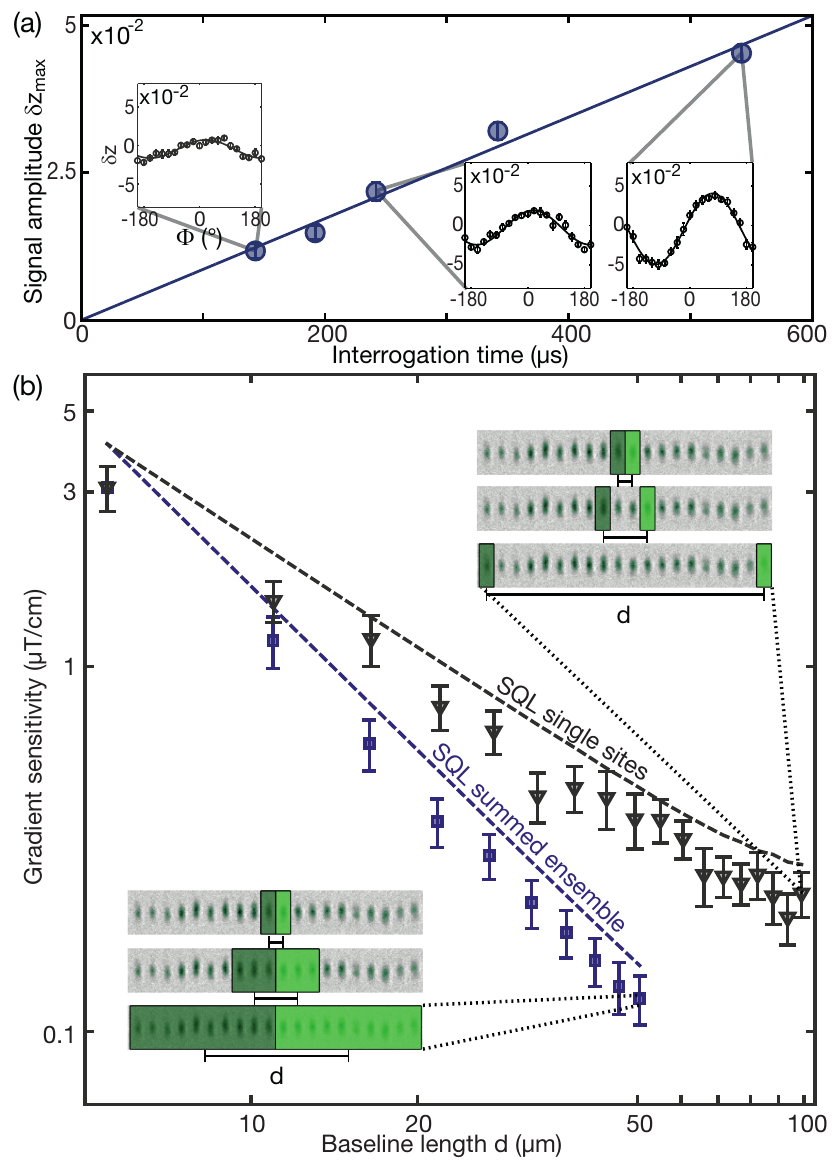}
\caption{(color online). {\bf Magnetic field gradiometry with an array of condensates.} (a) A magnetic field gradient is deduced from the fringe amplitude of $\delta z$ (insets), which grows linearly with interrogation time (blue circles and linear fit). We find $\partial{B_z}/ \partial{x} = 19.6(6)$\,pT/\textmu m.  (b) Gradiometric sensitivity for $t_{\text{int}} = 342$\,\textmu s. For increasing baseline length d between single wells, we find the expected linear gain of gradient sensitivity (triangles) below the classical limit  (SQL, upper dashed line).  The sensitivity can be further increased (squares) by summing the populations of lattice sites as indicated in the inset, and beats the corresponding standard quantum limit for the summed ensemble (lower dashed line). Error bars are the 1\,s.d. confidence from a resampling method.}
\label{Fig4}
\end{figure}
The single shot magnetic field sensitivity using this resource can be extracted by a differential measurement, which rejects fluctuations of homogeneous fields to first order. A magnetic field difference $\delta B$ between the left and right part of the ensemble translates into a differential phase $\delta \varphi$ of the Ramsey fringes  (Fig. 3(a)). For a fixed pulse phase, this shows up as $\delta z$, the difference of the corresponding population imbalances.  The optimal working point for estimating magnetic fields is close to the zero-crossings of the Ramsey fringes, where $\delta z$ is maximal. At this point, the difference in the magnetic field can be deduced as $\delta B \approx (\delta z)_{\text{max}}/2\pi t_{\text{int}} \mathcal{S} \mathcal{V}$ with the interrogation time $t_{\text{int}} = t_{\text{hold}} + 2t_{\pi}$. The single shot magnetic field sensitivity around the optimal working point follows from error propagation in the expression for $\delta B$ using the measured values for Var($\delta z$)  \cite{SuppInfo}. We find quantum-enhanced sensitivity up to interrogation times of 342\,\textmu s (see Fig. 3(b)). For longer times, quantum enhancement is lost due to fluctuations of the magnetic field which translate into a significant reduction of the mean Ramsey contrast. We do not observe a decrease in single-shot visibility, indicating that no coherence is lost on these timescales. For $t_{\text{int}} = 342$\,\textmu s and the full ensemble, we find a quantum-enhanced single shot sensitivity for static magnetic fields of  $310(47)$\,pT compared to the shot noise limit of 382\,pT for a perfect classical device (same atom number, no detection noise, and $\mathcal{V} = 1$). With our current experimental duty cycle (36\,s production, 342\,\textmu s interrogation),  we realized a sub-shot noise sensitivity of $1.86(28)$\,nT/$\sqrt{\text{Hz}}$ for static magnetic fields. The performance of our magnetometer is competitive with state-of-the-art devices with comparably small probe volume~\cite{Budker2013BOOK}, such as micro-SQUIDs or nitrogen vacancy centers (see \cite{SuppInfo} for an overview). The ultimate physical limitation is the residual nonlinearity of the employed states, which implies that further improvement of the sensitivity by at least two orders of magnitude can be achieved by increasing the interrogation time. Thus, for an interrogation time of 250\,ms~\cite{VengalattorePRL2007} and assuming a realistic cycle time of 5\,s for an all-optical BEC apparatus, a sensitivity of  $\sim$ 1\,pT/$\sqrt{\text{Hz}}$  in a probe volume of just 90\,\textmu m$^3$ is feasible.\\ 
 Our array of BECs is ideally suited for gradiometric measurements. The sensitivity for magnetic field gradients depends on interrogation time, the distance between the detectors (baseline length), and the noise of the magnetic field detection. We find the expected linear dependence of the signal amplitude  $(\delta z)_{\text{max}}$ on interrogation time, as shown in  Fig. 4(a).  
The gradiometric sensitivity for the specific interrogation time of 342\,\textmu s as a function of baseline length is shown in Fig. 4(b). Using single wells with varying distance, we find the expected linear gain in gradient sensitivity with baseline length (triangles) and observe quantum enhancement beyond the respective classical limits (dashed line).\\
The sensitivity of the gradiometer can be further improved to $12(2)$\,pT/\textmu m by summing adjacent lattice sites, exploiting the scalability of the squeezed state, and leading to a mean baseline length of up to 50\,\textmu m (Fig. 4(b), squares). Specifically for our experiment, we measure a gradient of $\partial{B_z}/ \partial{x} = 19.6(6)$\,pT/\textmu m with quantum enhancement of up to 24\,\%, as expected from the independently determined 2.4\,dB of squeezing (see Fig. 2(b) at $\alpha = 90$\textdegree, here without subtraction of detection noise).\\
In conclusion, we demonstrated the scalability of squeezed ensembles in an optical lattice, directly applicable to high-precision atom interferometry with ultracold clouds. 
We show that swapping of squeezed states is experimentally feasible and allows both for control of nonlinear interaction and the field sensitivity. Both advantages are explicitly demonstrated with quantum-enhanced Ramsey magnetometry, achieving high sensitivity and spatial resolution. The flexibility of state swapping with large squeezed ensembles offers prospects for improved tests of general relativity or the detection of gravitational waves with atom interferometers employing motional degrees of freedom~\cite{GrahamPRL2013, DimopoulosPRL2007}, controlled by Raman beam splitters. 
From a more fundamental perspective, the system of several independent ensembles entangled in the internal degrees of freedom combined with adjustable tunnel coupling is a perfect starting point to study the spread of quantum correlations~\cite{CheneauNATURE2012, LangenNATUREPHYS2013} in the continuous variable limit and the role of entanglement in quantum phase transitions~\cite{OsterlohNATURE2002, OsbornePRA2002}. \\

We thank I. Stroescu and  J. Schulz for technical help and discussions.
This work was supported by the Heidelberg Center for Quantum Dynamics and 
the European Commission small or medium-scale focused research project QIBEC (Quantum Interferometry with Bose-Einstein condensates, Contract Nr. 284584).
W.M. acknowledges support by the Studienstiftung des deutschen Volkes.
D.B.H. acknowledges support from the Alexander von Humboldt foundation.


\bibliographystyle{APS}

\newpage
\clearpage
\onecolumngrid
\renewcommand{\figurename}{Supp.~Fig.}
\setcounter{figure}{0}   
\begin{center} 
\section*{Supplementary Material: Scalable Spin Squeezing for Quantum-Enhanced Magnetometry with Bose-Einstein Condensates}

W. Muessel$^{*}$,  H. Strobel, D. Linnemann, D. B. Hume \& M. K. Oberthaler

\vspace{0.5cm}

{\it $^1$Kirchhoff-Institut f\"ur Physik, Universit\"at Heidelberg, Im Neuenheimer Feld 227, 69120 Heidelberg, Germany.}

\end{center}

\subsection{Comparison of magnetic field sensitivity to state-of-the-art magnetometers}

 \begin{figure}[ht]
\includegraphics[scale=1.2]{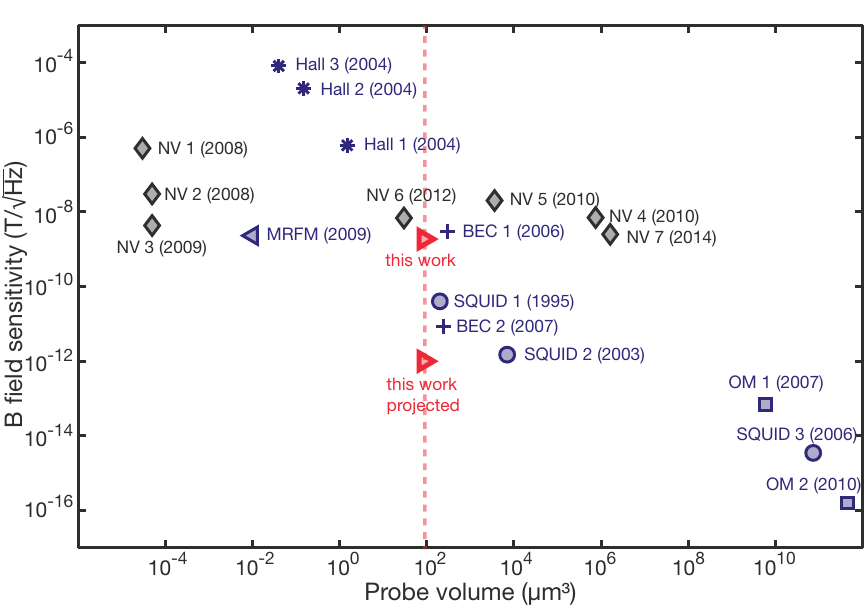}
\caption{{\bf Sensitivity of state-of-the-art magnetometers vs. probe volume.} The sensitivity of our magnetometer is on a competitive level with state-of-the-art magnetometers of similar probe volumes. For comparison, we show the sensitivities for  NV centers \cite{MazeNATURE2008} (NV1,2) and \cite{ BalasubramanianNATMAT2009, SteinertRSI2010, AcostaAPL2010, PhamPRB2012, JensenPRL2014} (NV 3-7), Hall probes 1-3 \cite{SandhuME2004}, optical vapor cell magnetometers  (OM 1 and 2) \cite{ShahNatPhot2007, DangAPL2010}, SQUIDS 1-3 \cite{KirtleyAPL1995, BaudenbacherAPL2003, FaleyJoP2006}, Bose-Einstein condensates (BEC1 and 2) \cite{WildermuthAPL2006, VengalattorePRL2007} and magnetic resonance force microscopy (MRFM) \cite{MaminNanoLett2009}. It is important to note that our sensitivity is reached for static fields, while most sensors in this figure can only reach the given sensitivities for AC fields and perform worse at the DC level.}
\label{SuppFig1}
\end{figure}
\subsection{Squeezing generation on single lattice sites}
All atoms on an individual lattice site occupy in good approximation a single spatial mode, such that only internal dynamics takes place. The states  $|\text{a} \rangle$ and  $|\text{b} \rangle$ form an effective two-level system for the $N$ particles, which can be described as a Schwinger pseudospin with length $\langle \hat{J} \rangle = N/2$, where $\hat{J}_z = (\hat{N_2} - \hat{N_1}) / 2$ and $\hat{J_x}$ and $\hat{J_y}$ are the corresponding coherences. In this description, the nonlinear interaction between the atoms,  introduced by an interspecies Feshbach resonance at 9.1 G,  leads to the term $\chi \hat{J}_z ^2$ in the Hamiltonian. Here, $\chi$ parametrizes the interaction strength. Including a detuning $\delta$ with respect  to the atomic resonance frequency, the time evolution of the system is governed by the Hamiltonian
\begin{equation}
\mathcal{H} = \chi \hat{J}_z ^2 + \delta  \hat{J}_z.
\end{equation}
This Hamiltonian is known as the one-axis twisting Hamiltonian \cite{KitagawaPRA1993} and leads to a redistribution of uncertainties. Our initial state is an ensemble of $N$ independently prepared atoms in the same superposition state, i.e. a coherent spin state of the total ensemble. The nonlinear evolution leads to reduction of quantum uncertainty along a certain axis and corresponding increase in orthogonal direction. In our experimental system, squeezing is limited by two-body relaxation loss from the excited state $|\text{b} \rangle$ (timescale 200\,ms) and loss due to the proximity of the Feshbach resonance (combined timescale 110\,ms), limiting the theoretically attainable spin squeezing to $\approx$ -9 dB. 
 \begin{figure}[th]
\includegraphics{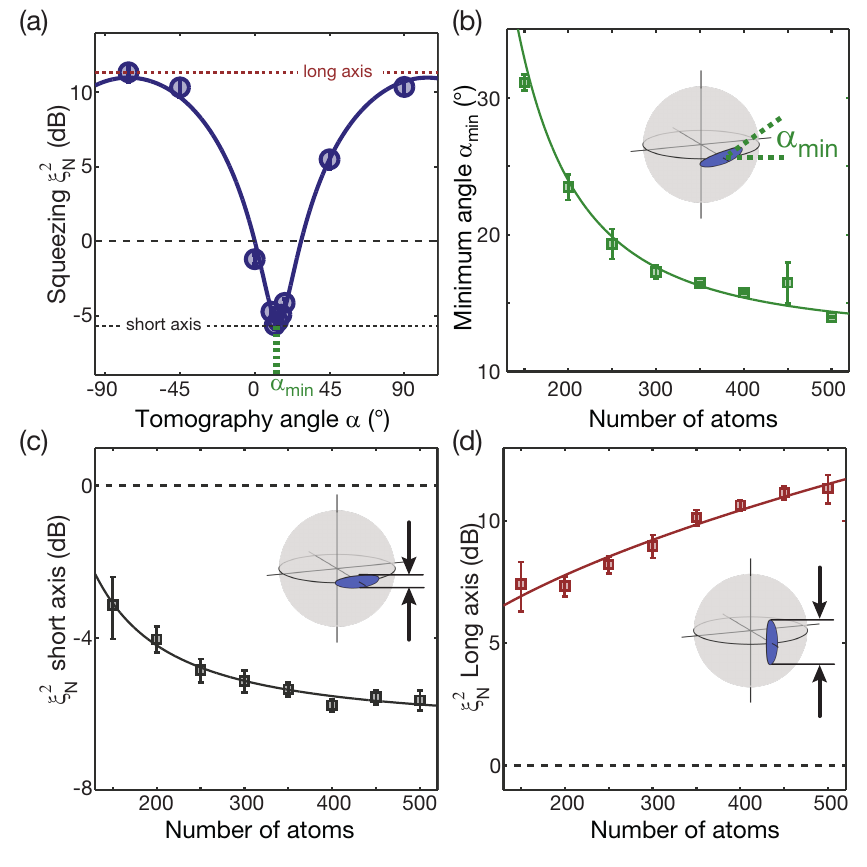}
\caption{{\bf Atom number dependence of final state on a single lattice site.} (a) After 20 ms of nonlinear evolution, we find, on each lattice site, squeezed states which exhibit a characteristic sinusoidal change of $\xi_{\text{N}}^2$ with rotation angle.  (b) The optimal rotation angle $\alpha_{\text{min}}$ depends only weakly on atom number for $N>300$. (c) The minimal number squeezing $\xi_{\text{N}}^2$  also shows only weak dependence on atom number. (d) Fluctuations along the axis of maximal uncertainty increase with atom number due to the increased nonlinear interaction strength. All solid lines are guides to the eye.}
\label{SuppFig2}
\end{figure}
\subsubsection*{N dependence of the Hamiltonian}
In our system, both nonlinear interaction $\chi(N)$ and detuning $\delta(N)$  depend on the total number of atoms. From independent measurements, we find a dependence of  $\delta(N) \approx \delta_0 + 1\text{Hz}/\text{(40 atoms)}$ and $\chi(N) \propto 1/\sqrt{N}$, with $\chi(500\text{ atoms}) = 2\pi\times 0.064$\, Hz, leading to an effective nonlinear interaction energy of  $N\chi \approx 2\pi\times 30$\, Hz. This leads to an atom number dependent squeezing factor and optimal readout angle. Supp.~Fig. 2 shows the dependence of these parameters on atom number  after 20 ms of nonlinear one-axis-twisting evolution for single lattice sites. We find weak dependence of both minimal number squeezing and optimal angle for atom numbers $N_{\text{tot}}>300$.
\subsection{Experimental sequences}
We prepare our 1D array of  Bose-Einstein condensates using a far off-resonant dipole trap at 1030\,nm and a standing wave potential generated by interference of two beams from a single laser source at 820\,nm,  yielding a lattice spacing of 5.5\,\textmu m. The resulting trap frequencies of the individual lattice sites are $\omega_l = 660$\,Hz in lattice direction and $\omega_t =130$\,Hz in transversal direction.
All atoms are initially condensed in  $|\text{c} \rangle = |F=1, m_F=-1\rangle $. After the cooling procedure, the bias field is ramped to 9.12 G close to the interspecies Feshbach resonance at 9.1 G and all atoms are transferred to the $|\text{a} \rangle = |1, 1\rangle$  state by a rapid adiabatic passage.
 The total cycle time for generating and probing the condensates is 36\,s.\\
Coupling between  $|\text{a} \rangle$ and  $|\text{b} \rangle = |2, -1\rangle$ is provided by a two-photon transition with combined radio frequency and microwave coupling 200\,kHz  red-detuned to the respective transitions to the $|2, 0\rangle $ level. The resulting two-photon Rabi frequency is 310\,Hz, calibrated by Rabi flopping. Control of phase, amplitude and frequency of the coupling is done via the arbitrary waveform generator that produces the radio frequency signal. During the long experimental timescales (several days), the resonance condition for the pulses is ensured by interleaved Ramsey experiments on the two-photon transition, which is also sensitive on the AC Zeeman shift of $\approx 120$\,Hz due to the off-resonant microwave and  $\approx 70$\,Hz due to off-resonant RF radiation during two-photon coupling. Transfer from  $|\text{b} \rangle $  to the level  $|\text{c} \rangle$ is realized by resonant one-photon microwave coupling with a Rabi frequency of 7 kHz.\\
\subsubsection*{Pulse sequence for one-axis twisting}
 For the generation of squeezed states via one-axis twisting, we initially prepare all atoms in an equal superposition between $|\text{a} \rangle $  and $|\text{b} \rangle $ using a $\pi$/2-pulse of the two-photon microwave and radio frequency coupling. After 10 ms of nonlinear evolution, a spin-echo $\pi$-pulse is applied. This spin-echo pulse has a relative phase of $\phi = 3\pi/2$ with respect to the initial $\pi$/2-pulse and reduces the susceptibility of the final state to technical detuning fluctuations (detailed below). After a second period of nonlinear evolution, tomographic readout with rotation  angle $\alpha$ is performed by applying a two-photon pulse with variable length. We choose a rotation phase of $\phi = \pi/2$ for $\alpha>0$ (rotation of axis with maximal fluctuations towards equator) and $\phi = 3/2 \pi$ for $\alpha<0$  to minimize the pulse length for readout. 
\subsubsection*{Pulse sequence for Ramsey interferometry}
In the quantum-enhanced Ramsey scheme, we first employ the one-axis twisting scenario with 20 ms of nonlinear evolution as described above. Here, the final tomography pulse rotates the squeezed state to its phase-squeezed axis, corresponding to a 75.5\textdegree\, rotation with $\phi = 3\pi/2$ (Supp.~Fig. 3(b)). For state swapping and interferometry, a one-photon microwave $\pi$-pulse transfers the population from  $|\text{b} \rangle $ to $|\text{c} \rangle $. After variable hold time for phase evolution, a second one-photon microwave $\pi$-pulse transfers the population back to the level $|\text{b} \rangle $. For the squeezing tomography in Fig. 2b, the final two-photon pulse is performed with fixed phase ($\phi = \pi$/2 or 3$\pi$/2 as for the tomography readout) and variable pulse length. For Ramsey fringe measurements (Fig. 2c), we apply a $\pi/2$-pulse with variable phase.
\subsubsection*{Readout of the atomic populations}
For population readout after the experimental sequence, we apply a $\pi$-pulse to transfer the population from  $|\text{b} \rangle $  to  $|\text{c} \rangle $ in order to avoid further dipole relaxation loss in the F=2 manifold. Subsequently, the bias field is ramped down to $\approx$~1 G . The two components are spatially separated via a Stern-Gerlach pulse and the trap is switched off for a short time-of-flight of 1.2 ms. Readout of the individual state populations is done with resonant absorption imaging on the D2 line with a spatial resolution of $\approx$ 1\,\textmu m, clearly resolving the single lattice sites. The resulting detection noise for each cloud is $\Delta_{\text{Det}} \approx \pm$4 atoms \cite{Muessel2013}.  Symmetric detection is ensured by using $\pi$ polarized imaging light. 
 \begin{figure}[th]
\includegraphics{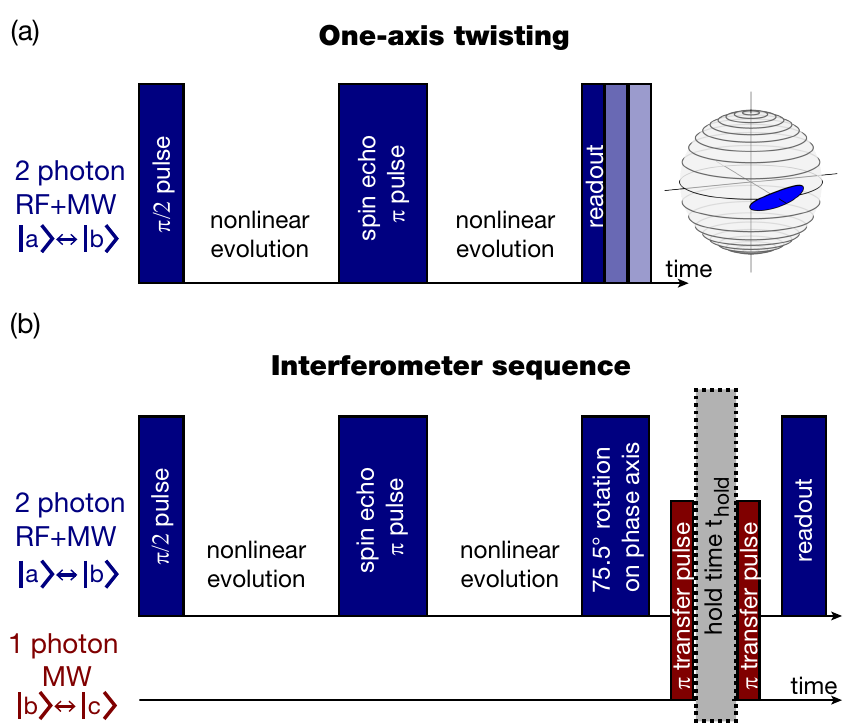}
\caption{{\bf Pulse sequences.} (a) For the generation of squeezed states via one-axis twisting, the intial state is prepared with a $\pi$/2-pulse using two-photon microwave and radio frequency coupling with $ \Omega \approx 2\pi\times310$\,Hz. After subsequent nonlinear evolution of 10 ms  a spin-echo $\pi$-pulse is applied, followed by a second period of 10 ms nonlinear evolution. The rotation for state characterization is done by varying the length of the final readout pulse. (b) For the interferometric sequence, in analogy to a) a squeezed state is produced and rotated by 75.5\textdegree,  leading to a phase-squeezed state. One-photon microwave $\pi$-pulses transfer the population from $|\text{b} \rangle $ to $|\text{c} \rangle $  and, after subsequent hold time for phase evolution, back to $|\text{b} \rangle $. Phase-controlled readout is done using two-photon coupling between $|\text{a} \rangle $ and $|\text{b} \rangle $. }
\label{SuppFig3}
\end{figure}
\subsection{Extraction of squeezing for large ensembles}
\subsubsection*{Direct analysis}
We calculate the number squeezing factor $\xi^2_{\text{N}} = \frac{2\text{Var}( J_z)}{J}$ from the variance of the population difference $\Delta^2_- = \text{Var} (N_{\uparrow}-N_{\downarrow})$ and the total atom number $N_{\text{tot}}$. This yields  $\xi^2_{\text{N}} =\frac{\Delta^2_- - \Delta^2_{\text{Det}}}{4p(1-p)N_{\text{tot}}}$ correcting for the mean imbalance with the binomial factor $p = 1/2 + \frac{\langle(N_{\uparrow}-N_{\downarrow})\rangle}{2N_{\text{tot}}}$ and the independently characterized detection noise $ \Delta^2_{\text{Det}}$ of our imaging system \cite{Muessel2013}. The metrological spin squeezing parameter  $\xi^2_{\text{R}} = \frac{\xi^2_{\text{N}}}{\mathcal{V}^2}$ quantifies the attainable metrological gain and also takes into account technical imperfections that lead to a reduced visibility $\mathcal{V}$ of the Ramsey fringe.
\subsubsection*{Relative analysis}
With the parallel production of spin squeezed states, a self-referenced analysis of the squeezing allows for a rejection of common mode fluctuations. This relative analysis is directly related to the sensitivity attainable in gradiometry.
Dividing the system into two parts with respective population imbalances  $ z_1 =\frac{ N_{1-}}{N_{1+}}$  and  $ z_2 =\frac{ N_{2-}}{N_{2+}}$, noise suppression is found in the imbalance difference $\delta z = z_1 - z_2$ . Here,  $N_{1+}$ and $N_{2+}$  are the total populations of the two samples and  $N_{1-}$ and $N_{2-}$  are the population differences. 
For classical states and two independent samples, we expect fluctuations of 
\begin{equation}
\text{Var} (\delta z)_{\text{class}} =  \left[ \text{Var}\left(\frac{ N_{1-}}{N_{1+}}\right) + \text{Var} \left(\frac{ N_{2-}}{N_{2+}}\right)\right]_{\text{class}} =\frac{c_1}{N_{1+}}+\frac{c_2}{N_{2+}},
\end{equation}
 using the binomial factors $c_1 = 4p_1(1-p_1)$ and $c_2= 4p_2(1-p_2)$  accounting for the finite individual imbalances. Note that in the case of equal sample sizes and $\langle z_1 \rangle= \langle z_2 \rangle= 0$, this simplifies to $\text{Var} (\delta z)_{\text{class}}  = 4/N_{\text{tot}}$ with the total number of atoms $N_{\text{tot}} = {N_{1+}}+{N_{2+}}$ .
 
   The relative squeezing factor is the ratio between the experimentally measured variance and the classical limit, yielding
\begin{equation}
\xi^2_{\text{rel}} = \frac{\text{Var} (\delta z)_{\text{exp}}}{\text{Var} (\delta z)_{\text{class}}} \approx\frac{ N_{\text{tot}}}{4} \cdot \text{Var} (\delta z)_{\text{exp}}.
\end{equation}
The relationship between relative squeezing  $\xi^2_{\text{rel}}$ and the number squeezing parameter  $\xi^2_{\text{N}}$ in the absence of technical (i.e. common mode) fluctuations is given by
\begin{equation}
\xi^2_{\text{rel}} = \frac{N_{2+}}{N_{\text{tot}}} \cdot   \xi^2_{\text{N1}} +   \frac{N_{1+}}{N_{\text{tot}}} \cdot \xi^2_{\text{N2}}, 
\end{equation}
using the individual number squeezing parameters  $\xi^2_{\text{N1}}$ and  $\xi^2_{\text{N2}}$ and assuming $c_1 = c_2$. Thus for similar atom numbers or identical individual number squeezing parameters, $\xi^2_{\text{rel}}$ is equivalent to the number squeezing parameter 
\begin{equation}
\xi^2_{\text{Ntot}}= \frac{N_{1+}}{N_{\text{tot}}} \cdot  \xi^2_{\text{N1}} +   \frac{N_{2+}}{N_{\text{tot}}} \cdot \xi^2_{\text{N2}} , 
\end{equation}
 of the total cloud. In our case, both of these conditions are well fulfilled, allowing direct comparison between the two parameters, as $\xi^2_{\text{Ntot}} \approx \xi^2_{\text{rel}}$.

\subsubsection*{Differential phase estimation in the magnetometry sequence}
For equal visibilities $\mathcal{V}$ in two samples,  $\delta z$  is related to the phase difference $\delta \phi$ by 
\begin{eqnarray}
\delta z & = \mathcal{V}\left( \sin\left(\phi_{\text{left}}\right) - \sin\left(\phi_{\text{left}} + \delta \phi\right)\right) \\
& = -2 \mathcal{V}\sin\left(\frac{\delta \phi}{2}\right)\cos\left(\frac{\delta \phi}{2}+\phi_{\text{left}}\right),
\end{eqnarray}
using the offset phase $\phi_{\text{left}}$ and $\phi_{\text{right}} = \phi_{\text{left}}+ \delta \phi$.
The corresponding difference in magnetic field can be estimated from the fringe amplitude $\delta z_{\text{max}} = 2 \mathcal{V}\sin\left(\frac{\delta \phi}{2}\right)$ using the magnetic field sensitivity $\mathcal{S}$ and the interrogation time $t_{\text{int}}$ as 
\begin{equation}
\delta B  = \frac{2 \arcsin\left( \frac{\delta z_{\text{max}}}{2 \mathcal{V}}\right)}{2\pi \mathcal{S} t_{\text{int}}}.
\label{FieldDiff}
\end{equation}
For small $\delta z_{\text{max}}$,   $\arcsin\left( \frac{\delta z_{\text{max}}}{2 \mathcal{V}}\right) \approx  \frac{\delta z_{\text{max}}}{2 \mathcal{V}}$. 
Error propagation of Eq. \ref{FieldDiff}  yields a sensitivity $\sigma_B$ close to the working point of
\begin{equation}
\sigma_B  = \frac{\text{Std}\left(\delta z_{\text{max}}\right)}{2\pi \mathcal{V}\mathcal{S} t_{\text{int}}}.
\label{FieldError}
\end{equation}
Here, the shot noise limited sensitivity is given by
\begin{equation}
\sigma_{\text{B SQL}}  = \frac{1}{2\pi \mathcal{V}\mathcal{S} t_{\text{int}}\sqrt{N_{\text{tot}}}}.
\end{equation}
The sensitivity of our device, even though being a proof-of-principle demonstration, is already on a competitive level in comparison to other state-of-the-art magnetometry techniques with comparable probe volume (see Supp.~Fig. 1). The projected sensitivity of 1\,pT/$\sqrt{\text{Hz}}$ exceeds that of all current techniques with comparable spatial resolution.
\subsection{Magnetic field noise}
The main source of technical noise in our system are shot-to-shot fluctuations of the magnetic bias field at 9.12 G. This field is generated using a large pair of coils (edge length~$\approx$~1\,m) in Helmholtz configuration, and is actively stabilized using a fluxgate sensor in the vicinity of the experimental chamber. The dominant AC component at the line frequency of 50 Hz is compensated by a feed forward technique. Long-term drifts due to temperature or humidity changes are corrected using interleaved Ramsey measurements between $|\text{a} \rangle$ and  $|\text{b} \rangle$.\\
 This stabilization reduces the shot-to-shot fluctuations of the field to $\approx$ 30\,\textmu G, which we determine from the scatter of repeated Ramsey measurements on the magnetically sensitive transition between the  $|1, 1\rangle$  and the  $|2, 0\rangle$ level. Additional small long-term drifts lead to an effective long-term stability of $\approx$ 45\,\textmu G for a typical timescale of one weekend. A technical noise analysis of the prepared squeezed state is consistent with this value (below).  
 \subsubsection*{Effects on squeezing generation} 
 \begin{figure}[tp]
\includegraphics{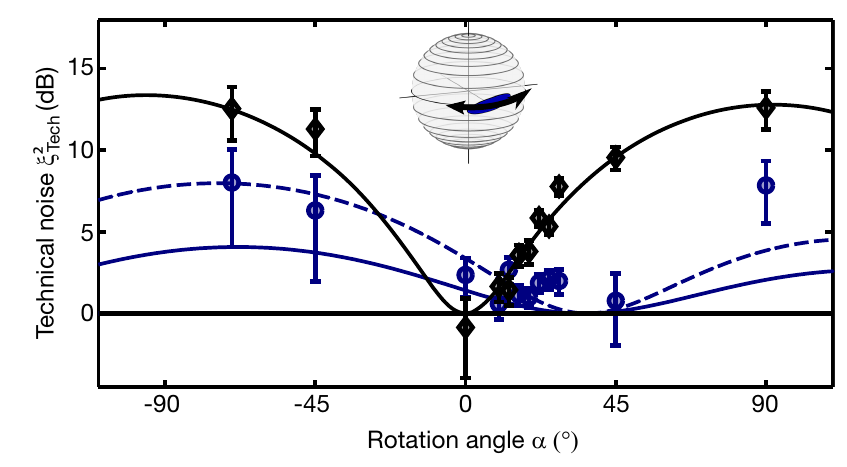}
\caption{{\bf Technical noise tomography for 15 ms of nonlinear one-axis twisting evolution.} For different readout rotations, the contribution of technical noise to the fluctuations (given for $10^4$ particles in units of coherent state uncertainty) is deduced from the quadratic scaling contribution to the variance of population differences with particle number. After 15 ms of nonlinear one-axis twisting evolution without spin-echo pulse (black diamonds), we find that all technical contributions are in phase direction (90\textdegree\,  rotation angle) and compatible with our independently measured bias field fluctuations of 45\,\textmu G (solid line: numerical simulation). A spin-echo pulse in the middle of the time evolution significantly suppresses the technical fluctuations (blue circles). The remaining noise contribution  can be explained by imperfections of the spin-echo pulse due to the presence of nonlinear interaction (simulation: solid blue line) and additional level shifts (e.g. fluctuation of AC Zeeman shift due to microwave and radio frequency) $\sigma_P =1.3$\,Hz (simulation: dashed line). Error bars are statistical 1 s.d. errors obtained from a resampling analysis.}
\label{SuppFig4}
\end{figure} 
Technical fluctuations lead to a linear dependence of $\xi_{\text{N}}^2$ versus atom number, corresponding to a quadratic component of the variance. This limits number squeezing for large sample sizes, whereas the relative analysis, which is insensitive to common mode fluctuations in first order, is unaffected. In the following section, we will explain the influence of technical fluctuations on our squeezing procedure.\\ 
 The transition between $|\text{a} \rangle$ and $|\text{b} \rangle$ is only quadratically sensitive magnetic field changes, yielding a sensitivity of 10\,Hz/mG at our bias field of 9.12 G. The shot-to-shot fluctuations of the field thus translate into detuning fluctuations of $\text{Std}(\delta) = 0.3$\,Hz.\\
   For the one-axis twisting Hamiltonian, the detuning fluctuations directly translate into increased phase fluctuations, which are transferred to number fluctuations by applying a tomography rotation.\\
We  tomographically investigate the influence of technical fluctuations by analyzing the mean number squeezing for different summing regions over the lattice.  The technical fluctuations contribute quadratically to the variance of the population difference $\Delta_- \propto \beta^2 N_{\text{tot}}^2$, and  thus can be quantified by $\beta^2$, which is deduced from a quadratic fit. Atom number dependent effects on the individual sites are minimized by averaging all possible combinations including all sites in the BEC array leading to a chosen mean atom number.\\
Supp.~Fig. 4 shows the technical noise contribution to the observed fluctuations of a state with $10^4$ atoms obtained after 15\,ms of one-axis twisting evolution. For nonlinear evolution without a spin-echo pulse, we find only phase fluctuations (black diamonds). The solid line  is obtained from numerical simulations with shot-to-shot detuning fluctuations of $\text{Std}(\delta) = 0.45$\,Hz given by the magnetic field stability. We find that a spin-echo pulse in the middle of the evolution significantly reduces the phase-noise contribution (blue circles). The largest part of the remaining noise is explained by the imperfect spin-echo pulse, which, due to the nonlinear interaction, is effectively reduced by an angle of $\approx$ 9\textdegree\,(numerical simulation: solid blue line). This translates a small fraction of the phase fluctuations into imbalance fluctuations after the spin-echo pulse. These, however, are strongly amplified during the second half of the nonlinear evolution. Additional detuning fluctuations of 1.5 Hz during the two-photon pulses, e.g. due to slight changes in the AC Zeeman shift, explains the additional contribution to the technical noise (dashed blue line). In principle, perfect compensation with the spin-echo pulse can be achieved by increasing the Rabi coupling strength or adjusting the pulse durations.\\ 
  \begin{figure}[tp]
\includegraphics{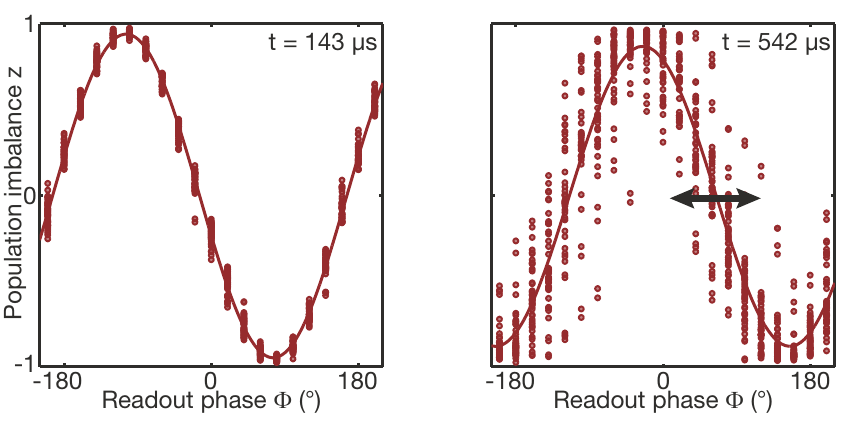}
\caption{{\bf Effects of technical noise on interferometric sensitivity.} The left panel shows the population imbalance of single shots  for varying readout phase in a Ramsey sequence after a hold time of 1\,\textmu s, which features a sinusoidal variation with a visibility of 95\% (solid line: sinusoidal fit). Shot-to-shot fluctuations of the magnetic bias field translate into different phase evolutions. This can be seen from the scatter in population imbalances for a Ramsey fringe with a hold time of 400\,\textmu s. This shift of the offset phase leads to shifts of the working point of the interferometer and reduces the mean visibility, which is used for calculation of the sensitivity. The single shot visibility remains high, indicating that this effect is solely  due to dephasing.}
\label{SuppFig5}
\end{figure} 
 \subsubsection*{Effects on interferometric sensitivity}
 Employing gradiometric measurements of the magnetic field, we can characterize the sensitivity of our magnetometer even in the presence of fluctuating homogeneous fields. These are cancelled to first order since in our case the respective phases of the Ramsey fringes, which are given by $\phi = 2\pi \mathcal{S} B t_{\text{int}}$ (see Supp.~Fig. 5), vary together. This approach is limited to  interrogation times where the accumulated phase fluctuations are smaller than $\pi$. It is important to note that during the interferometry sequence the shot-to-shot fluctuations of the offset field translate 140 times stronger into detuning fluctuations than during the generation procedure of the squeezed states, as the levels $|\text{a} \rangle$ and $|\text{c} \rangle$ are linearly sensitive to magnetic field changes.

\bibliographystyle{APS}

\begin{thebibliography}{10}
\expandafter\ifx\csname url\endcsname\relax
  \def\url#1{\texttt{#1}}\fi
\expandafter\ifx\csname urlprefix\endcsname\relax\def\urlprefix{URL }\fi
\providecommand{\bibinfo}[2]{#2}
\providecommand{\eprint}[2][]{\url{#2}}

 
 

\bibitem{CroninRMP2009}
\bibinfo{author}{A.~D. Cronin},
\bibinfo{author}{J. Schmiedmayer}, and \bibinfo{author}{D.~E. Pritchard},
\newblock { \bibinfo{journal}{Rev. Mod. Phys.}} 
\textbf{\bibinfo{volume}{81}},
\bibinfo{pages}{1051} 
(\bibinfo{year}{2009}).


\bibitem{GeigerNATCOMM2011}
\bibinfo{author}{R. Geiger} {\it et~al.},
\newblock { \bibinfo{journal}{Nat. Commun.}} 
\textbf{\bibinfo{volume}{2:474}},
\bibinfo{url}{doi: 10.1038/ncomms1479}
(\bibinfo{year}{2011}).

\bibitem{PetersMETROLOGIA2001}
\bibinfo{author}{A. Peters},
\bibinfo{author}{K.~Y. Cung}, and \bibinfo{author}{S. Chu},
\newblock {\bibinfo{journal}{Metrologia}} 
\textbf{\bibinfo{volume}{38}},
\bibinfo{pages}{25} 
(\bibinfo{year}{2001}).


\bibitem{GustavsonPRL1997}
\bibinfo{author}{T.~L. Gustavson},
\bibinfo{author}{P. Bouyer}, and \bibinfo{author}{M.~A. Kasevich},
\newblock {\bibinfo{journal}{Phys. Rev. Lett.}} 
\textbf{\bibinfo{volume}{78}},
\bibinfo{pages}{2046} 
(\bibinfo{year}{1997}).



\bibitem{WynandsMETROLOGIA2005}
\bibinfo{author}{R. Wynands} and
\bibinfo{author}{S. Weyers},
\newblock {\bibinfo{journal}{Metrologia}} 
\textbf{\bibinfo{volume}{42}},
\bibinfo{pages}{64} 
(\bibinfo{year}{2005}).



\bibitem{ItanoPRA1993}
\bibinfo{author}{W.~M. Itano},
\bibinfo{author}{J.~C. Bergquist},
\bibinfo{author}{J.~J. Bollinger}, 
\bibinfo{author}{J.~M. Gilligan}, 
\bibinfo{author}{D.~J. Heinzen}, 
\bibinfo{author}{F.~L. Moore}, 
\bibinfo{author}{M.~G. Raizen}, and 
\bibinfo{author}{D.~J. Wineland},
\newblock {\bibinfo{journal}{Phys. Rev. A}}
\textbf{\bibinfo{volume}{47}},
\bibinfo{pages}{3554}
(\bibinfo{year}{1993}).



\bibitem{WinelandPRA1994}
\bibinfo{author}{ D.~J. Wineland},
\bibinfo{author}{J.~J. Bollinger},
\bibinfo{author}{W.~M. Itano}, and \bibinfo{author}{D.~J. Heinzen},
\newblock {\bibinfo{journal}{Phys. Rev. A}}
\textbf{\bibinfo{volume}{50}},
\bibinfo{pages}{67}
(\bibinfo{year}{1994}).


\bibitem{GiovannettiSCIENCE2004}
\bibinfo{author}{V. Giovannetti},
\bibinfo{author}{S. Lloyd}, and \bibinfo{author}{L. Maccone},
\newblock {\bibinfo{journal}{Science}} 
\textbf{\bibinfo{volume}{306}},
\bibinfo{pages}{1330} 
(\bibinfo{year}{2004}).

\bibitem{LigoNatPhys2011}
\bibinfo{author}{The LIGO Scientific Collaboration.}
\newblock {\bibinfo{journal}{Nature Phys.}}
 \textbf{\bibinfo{volume}{7}},
 \bibinfo{pages}{962} (\bibinfo{year}{2011}).
 
 
\bibitem{EsteveNATURE2008}
\bibinfo{author}{J. Est\`eve}, 
\bibinfo{author}{C. Gross}, 
\bibinfo{author}{A. Weller}, 
\bibinfo{author}{S. Giovanazzi}, and 
\bibinfo{author}{M.~K. Oberthaler},
\newblock {\bibinfo{journal}{Nature}} \textbf{\bibinfo{volume}{455}},
 \bibinfo{pages}{1216} (\bibinfo{year}{2008}).
 
  
\bibitem{AppelPNAS2009}
\bibinfo{author}{J. Appel},
\bibinfo{author}{P.~J. Windpassinger},
\bibinfo{author}{D. Oblak},
\bibinfo{author}{U. B. Hoff},
\bibinfo{author}{N. Kj\ae rgaard}, and 
\bibinfo{author}{E.~S. Polzik},
\newblock {\bibinfo{journal}{Proc. Nat. Acad. Sci. USA}} 
\textbf{\bibinfo{volume}{106}},
\bibinfo{pages}{10960} (\bibinfo{year}{2009}).
 
\bibitem{GrossNATURE2010}
\bibinfo{author}{C. Gross}, 
\bibinfo{author}{T. Zibold}, 
\bibinfo{author}{E. Nicklas}, 
\bibinfo{author}{J. Est\`eve}, and 
\bibinfo{author}{M.~K. Oberthaler},
\newblock {\bibinfo{journal}{Nature}} \textbf{\bibinfo{volume}{464}},
 \bibinfo{pages}{1165} (\bibinfo{year}{2010}).

\bibitem{RiedelNATURE2010}
\bibinfo{author}{M.~F. Riedel},
\bibinfo{author}{P. B{\"o}hi},
\bibinfo{author}{Y. Li}, 
\bibinfo{author}{T.~W. H{\"a}nsch},
\bibinfo{author}{A. Sinatra}, and
\bibinfo{author}{P. Treutlein},
\newblock {\bibinfo{journal}{Nature}} \textbf{\bibinfo{volume}{464}},
 \bibinfo{pages}{1170} (\bibinfo{year}{2010}).

  
 \bibitem{LerouxPRL2010-CS}
\bibinfo{author}{I.~D. Leroux}, 
\bibinfo{author}{M.~H. Schleier-Smith}, and 
\bibinfo{author}{V. Vuleti\'c},
\newblock {\bibinfo{journal}{Phys. Rev. Lett.}} \textbf{\bibinfo{volume}{104}},
 \bibinfo{pages}{073602} (\bibinfo{year}{2010}).
 
 
\bibitem{ChenPRL2011}
\bibinfo{author}{Z. Chen}, 
\bibinfo{author}{J.~G. Bohnet}, 
\bibinfo{author}{S.~R. Sankar}, 
\bibinfo{author}{J. Dai}, and 
\bibinfo{author}{J.~K. Thompson},
\newblock {\bibinfo{journal}{Phys. Rev. Lett.}}
 \textbf{\bibinfo{volume}{106}},
 \bibinfo{pages}{133601} (\bibinfo{year}{2011}).


\bibitem{SewellPRL2012}
\bibinfo{author}{R.~J. Sewell},
\bibinfo{author}{M. Koschorreck},
\bibinfo{author}{M. Napolitano},
\bibinfo{author}{B. Dubost},
\bibinfo{author}{N. Behbood}, and 
\bibinfo{author}{M.~W. Mitchell},
\newblock {\bibinfo{journal}{Phys. Rev. Lett.}}
 \textbf{\bibinfo{volume}{109}},
 \bibinfo{pages}{253605} (\bibinfo{year}{2012}).
 
\bibitem{BerradaNATCOMM2013}
\bibinfo{author}{T. Berrada},
\bibinfo{author}{S. van~Frank}, 
\bibinfo{author}{R. B{\"u}cker},
\bibinfo{author}{T. Schumm}, 
\bibinfo{author}{J.-F. Schaff}, and 
\bibinfo{author}{J. Schmiedmayer},
\newblock {\bibinfo{journal}{Nat. Commun.}} \textbf{\bibinfo{volume}{4:2077}},
 \bibinfo{url}{doi: 10.1038/ncomms3077} (\bibinfo{year}{2013}).


\bibitem{OckeloenPRL2013}
\bibinfo{author}{C.~F. Ockeloen}, 
\bibinfo{author}{R. Schmied}, 
\bibinfo{author}{M.~F. Riedel}, and 
\bibinfo{author}{P. Treutlein},
\newblock {\bibinfo{journal}{Phys. Rev. Lett.}}
 \textbf{\bibinfo{volume}{111}},
 \bibinfo{pages}{143001} (\bibinfo{year}{2013}).


\bibitem{LouchetChauvetNJP2010}
\bibinfo{author}{A. Louchet-Chauvet}, 
\bibinfo{author}{J. Appel}, 
\bibinfo{author}{J.~J. Renema},
\bibinfo{author}{D. Oblak}, 
\bibinfo{author}{N. Kjaergaard}, and 
\bibinfo{author}{E.~S. Polzik},
\newblock {\bibinfo{journal}{New J. Phys.}} 
\textbf{\bibinfo{volume}{12}},
\bibinfo{pages}{065032} 
(\bibinfo{year}{2010}).

\bibitem{LerouxPRL2010}
\bibinfo{author}{I.~D. Leroux}, 
\bibinfo{author}{M.~H. Schleier-Smith}, and 
\bibinfo{author}{V. Vuleti\'c},
\newblock {\bibinfo{journal}{Phys. Rev. Lett.}} 
\textbf{\bibinfo{volume}{104}},
 \bibinfo{pages}{250801}
  (\bibinfo{year}{2010}).


\bibitem{KitagawaPRA1993}
\bibinfo{author}{M. Kitagawa} and 
\bibinfo{author}{M. Ueda},
\newblock {\bibinfo{journal}{Phys. Rev. A}} 
\textbf{\bibinfo{volume}{47}},
\bibinfo{pages}{5138} 
(\bibinfo{year}{1993}).


\bibitem{SuppInfo}
\bibinfo{title}{See Supplemental Material at}
\bibinfo{url}{http://journals.aps.org/prl}
\bibinfo{title}{for further comparison with other magnetometry techniques, details on the experimental sequence and analysis methods.}

\bibitem{VengalattorePRL2007}
\bibinfo{author}{M. Vengalattore},
\bibinfo{author}{J.~M. Higbie},
\bibinfo{author}{S.~R. Leslie},
\bibinfo{author}{J. Guzman},
\bibinfo{author}{L.~E. Sadler}, and 
\bibinfo{author}{D.~M. Stamper-Kurn},
\newblock {\bibinfo{journal}{Phys. Rev. Lett.}}
\textbf{\bibinfo{volume}{98}},
\bibinfo{pages}{200801}
(\bibinfo{year}{2007}).


\bibitem{AignerSCIENCE2008}
\bibinfo{author}{S. Aigner},
\bibinfo{author}{L. Della Pietra},
\bibinfo{author}{Y. Japha},
\bibinfo{author}{O. Entin-Wohlman},
\bibinfo{author}{T. David},
\bibinfo{author}{R. Salem},
\bibinfo{author}{R. Folman}, and
\bibinfo{author}{J. Schmiedmayer},
\newblock {\bibinfo{journal}{Science}}
\textbf{\bibinfo{volume}{319}},
\bibinfo{pages}{1226}
(\bibinfo{year}{2008}).

\bibitem{Budker2013BOOK}
\bibinfo{author}{D. Budker} and \bibinfo{author}{D.~F.~J. Kimball},
\newblock \bibinfo{title}{{\it Optical magnetometry}}.
 (\bibinfo{publisher}{Cambridge}, \bibinfo{year}{2013}).
 
\bibitem{GrahamPRL2013}
\bibinfo{author}{P.~W. Graham},
\bibinfo{author}{J.~M. Hogan},
\bibinfo{author}{M.~A. Kasevich}, and 
\bibinfo{author}{S. Rajendran},
\newblock {\bibinfo{journal}{Phys. Rev. Lett.}} 
\textbf{\bibinfo{volume}{110}},
\bibinfo{pages}{171102} 
(\bibinfo{year}{2013}).


\bibitem{DimopoulosPRL2007}
\bibinfo{author}{S. Dimopoulos},
\bibinfo{author}{P.~W. Graham},
\bibinfo{author}{J.~M. Hogan}, and
\bibinfo{author}{M.~A. Kasevich}, 
\newblock {\bibinfo{journal}{Phys. Rev. Lett.}} 
\textbf{\bibinfo{volume}{98}},
\bibinfo{pages}{111102} 
(\bibinfo{year}{2007}).



 \bibitem{CheneauNATURE2012}
\bibinfo{author}{M. Cheneau} {\it et~al.},
\newblock {\bibinfo{journal}{Nature}}
\textbf{\bibinfo{volume}{481}},
\bibinfo{pages}{484}
(\bibinfo{year}{2012}).


\bibitem{LangenNATUREPHYS2013}
\bibinfo{author}{T. Langen},
\bibinfo{author}{R. Geiger},
\bibinfo{author}{M. Kuhnert},
\bibinfo{author}{B. Rauer}, and
\bibinfo{author}{J. Schmiedmayer}, 
\newblock {\bibinfo{journal}{Nature Phys.}}
\textbf{\bibinfo{volume}{9}},
\bibinfo{pages}{640}
(\bibinfo{year}{2013}).

\bibitem{OsterlohNATURE2002}
\bibinfo{author}{A. Osterloh},
\bibinfo{author}{L. Amico},
\bibinfo{author}{G. Falci}, and
\bibinfo{author}{R. Fazio},
\newblock {\bibinfo{journal}{Nature}}
\textbf{\bibinfo{volume}{416}},
\bibinfo{pages}{608}
(\bibinfo{year}{2002}).

\bibitem{OsbornePRA2002}
\bibinfo{author}{T.~J. Osborne} and
\bibinfo{author}{M.~A. Nielsen}, 
\newblock {\bibinfo{journal}{Phys. Rev. A}}
\textbf{\bibinfo{volume}{66}},
\bibinfo{pages}{032110}
(\bibinfo{year}{2002}).


\end{thebibliography}

\begin{thebibliography}{10}
\expandafter\ifx\csname url\endcsname\relax
  \def\url#1{\texttt{#1}}\fi
\expandafter\ifx\csname urlprefix\endcsname\relax\def\urlprefix{URL }\fi
\providecommand{\bibinfo}[2]{#2}
\providecommand{\eprint}[2][]{\url{#2}}

\bibitem{MazeNATURE2008}
\bibinfo{author}{J.~R. Maze}
{\it et~al.},
\newblock {\bibinfo{journal}{Nature}}
\textbf{\bibinfo{volume}{455}},
\bibinfo{pages}{644}
(\bibinfo{year}{2008}).

\bibitem{BalasubramanianNATMAT2009}
\bibinfo{author}{G. Balasubramanian}
{\it et~al.},
\newblock {\bibinfo{journal}{Nat. Materials}}
\textbf{\bibinfo{volume}{8}},
\bibinfo{pages}{383}
(\bibinfo{year}{2009}).

\bibitem{SteinertRSI2010}
\bibinfo{author}{S. Steinert},
\bibinfo{author}{F. Dolde},
\bibinfo{author}{P. Neumann},
\bibinfo{author}{A. Aird},
\bibinfo{author}{B. Naydenov},
\bibinfo{author}{G. Balasubramanian},
\bibinfo{author}{F. Jelezko}, and
\bibinfo{author}{J. Wrachtrup},
\newblock {\bibinfo{journal}{Rev. Sci. Instr.}}
\textbf{\bibinfo{volume}{81}},
\bibinfo{pages}{043705}
(\bibinfo{year}{2010}).

\bibitem{AcostaAPL2010}
\bibinfo{author}{V.~M. Acosta},
\bibinfo{author}{E. Bauch},
\bibinfo{author}{A. Jarmola},
\bibinfo{author}{L.~J. Zipp},
\bibinfo{author}{M. P. Ledbetter}, and
\bibinfo{author}{D. Budker},
\newblock {\bibinfo{journal}{Appl. Phys. Lett.}}
\textbf{\bibinfo{volume}{97}},
\bibinfo{pages}{174104}
(\bibinfo{year}{2010}).

\bibitem{PhamPRB2012}
\bibinfo{author}{L.~M. Pham},
\bibinfo{author}{N. Bar-Gill},
\bibinfo{author}{C. Belthangady},
\bibinfo{author}{D. Le Sage},
\bibinfo{author}{P. Cappellaro},
\bibinfo{author}{M. D. Lukin},
\bibinfo{author}{A. Yacoby}, and
\bibinfo{author}{R.~L. Walsworth},
\newblock {\bibinfo{journal}{Phys. Rev. B}}
\textbf{\bibinfo{volume}{86}},
\bibinfo{pages}{045214}
(\bibinfo{year}{2012}).

\bibitem{JensenPRL2014}
\bibinfo{author}{K. Jensen},
\bibinfo{author}{N. Leefer},
\bibinfo{author}{A. Jarmola},
\bibinfo{author}{Y. Dumeige},
\bibinfo{author}{V.~M. Acosta},
\bibinfo{author}{P. Kehayias},
\bibinfo{author}{B. Patton}, and
\bibinfo{author}{D. Budker},
\newblock {\bibinfo{journal}{Phys. Rev. Lett.}}
\textbf{\bibinfo{volume}{112}},
\bibinfo{pages}{160802}
(\bibinfo{year}{2014}).

\bibitem{SandhuME2004}
\bibinfo{author}{A. Sandhu},
\bibinfo{author}{A. Okamoto},
\bibinfo{author}{I. Shibasaki}, and
\bibinfo{author}{A. Oral},
\newblock {\bibinfo{journal}{Microelectronic engineering}}
\textbf{\bibinfo{volume}{73-74}},
\bibinfo{pages}{524-528}
(\bibinfo{year}{2004}).



 \bibitem{ShahNatPhot2007}
\bibinfo{author}{V. Shah},
\bibinfo{author}{S. Knappe},
\bibinfo{author}{P.~D. Schwindt}, and
\bibinfo{author}{J. Kitching}, 
\newblock {\bibinfo{journal}{Nat. Photon.}}
\textbf{\bibinfo{volume}{1}},
\bibinfo{pages}{649}
(\bibinfo{year}{2007}).

\bibitem{DangAPL2010}
\bibinfo{author}{H.~B. Dang},
\bibinfo{author}{A.~C. Maloof}, and
\bibinfo{author}{M.~V. Romalis},
\newblock {\bibinfo{journal}{Appl. Phys. Lett.}}
\textbf{\bibinfo{volume}{97}},
\bibinfo{pages}{151110}
(\bibinfo{year}{2010}).


\bibitem{KirtleyAPL1995}
\bibinfo{author}{J.~R. Kirtley},
\bibinfo{author}{M.~B. Ketchen},
\bibinfo{author}{K.~G. Stawiasz},
\bibinfo{author}{J.~Z. Sun},
\bibinfo{author}{W.~J. Gallagher},
\bibinfo{author}{S.~H. Blanton}, and
\bibinfo{author}{S.~J. Wind},
\newblock {\bibinfo{journal}{Appl. Phys. Lett.}}
\textbf{\bibinfo{volume}{66}},
\bibinfo{pages}{1138}
(\bibinfo{year}{1995}).


\bibitem{BaudenbacherAPL2003}
\bibinfo{author}{F. Baudenbacher},
\bibinfo{author}{L.~E. Fong},
\bibinfo{author}{J.~R. Holzer}, and
\bibinfo{author}{M. Radparvar},
\newblock {\bibinfo{journal}{Appl. Phys. Lett.}}
\textbf{\bibinfo{volume}{82}},
\bibinfo{pages}{3487}
(\bibinfo{year}{2003}).


\bibitem{FaleyJoP2006}
\bibinfo{author}{M.~I. Faley},
\bibinfo{author}{U. Poppe},
\bibinfo{author}{K. Urban},
\bibinfo{author}{D.~N. Paulson}, and
 \bibinfo{author}{R.~L. Fagaly},
\newblock {\bibinfo{journal}{Journal of Physics: Conference Series}}
\textbf{\bibinfo{volume}{43}},
\bibinfo{pages}{1199}
(\bibinfo{year}{2006}).



\bibitem{WildermuthAPL2006}
\bibinfo{author}{W. Wildermuth},
\bibinfo{author}{S. Hofferberth},
\bibinfo{author}{I. Lesanovsky},
\bibinfo{author}{S. Groth},
\bibinfo{author}{P. Kr{\"u}ger},
\bibinfo{author}{J. Schmiedmayer}, and
\bibinfo{author}{I. Bar-Joseph},
\newblock {\bibinfo{journal}{Appl. Phys. Lett.}}
\textbf{\bibinfo{volume}{88}},
\bibinfo{pages}{264103}
(\bibinfo{year}{2006}).

\bibitem{VengalattorePRL2007}
\bibinfo{author}{M. Vengalattore},
\bibinfo{author}{J.~M. Higbie},
\bibinfo{author}{S.~R. Leslie},
\bibinfo{author}{J. Guzman},
\bibinfo{author}{L.~E. Sadler}, and 
\bibinfo{author}{D.~M. Stamper-Kurn},
\newblock {\bibinfo{journal}{Phys. Rev. Lett.}}
\textbf{\bibinfo{volume}{98}},
\bibinfo{pages}{200801}
(\bibinfo{year}{2007}).


\bibitem{MaminNanoLett2009}
\bibinfo{author}{H.~J. Mamin},
\bibinfo{author}{T.~H. Oosterkamp},
\bibinfo{author}{M. Poggio},
\bibinfo{author}{C.~L. Degen},
\bibinfo{author}{C.~T. Rettner}, and
\bibinfo{author}{D. Rugar},
\newblock {\bibinfo{journal}{NANO LETTERS}}
\textbf{\bibinfo{volume}{9}},
\bibinfo{pages}{3020-3024}
(\bibinfo{year}{2009}).




\bibitem{KitagawaPRA1993}
\bibinfo{author}{M. Kitagawa} and 
\bibinfo{author}{M. Ueda},
\newblock {\bibinfo{journal}{Phys. Rev. A}} 
\textbf{\bibinfo{volume}{47}},
\bibinfo{pages}{5138} 
(\bibinfo{year}{1993}).


\bibitem{Muessel2013}
\bibinfo{author}{W. Muessel},
\bibinfo{author}{H. Strobel},
\bibinfo{author}{M. Joos},
\bibinfo{author}{E. Nicklas},
\bibinfo{author}{I. Stroescu},
\bibinfo{author}{J. Tomkovi\v{c}},
\bibinfo{author}{D.~B. Hume}, and
\bibinfo{author}{M.~K. Oberthaler},
\newblock {\bibinfo{journal}{Appl. Phys. B.}}
\textbf{\bibinfo{volume}{113}},
\bibinfo{pages}{69}
(\bibinfo{year}{2013}).

\end{thebibliography}

\end{document}